\documentstyle[prl,aps,epsf]{revtex}
\bibstyle{unsrt}
\tighten
\begin{document}
\draft
\preprint{LANL xxxx}
\def\be{\begin{equation}}
\def\ee{\end{equation}}
\def\bea{\begin{eqnarray}}
\def\eea{\end{eqnarray}}
\def\ben{\begin{eqnarray*}}
\def\een{\end{eqnarray*}}
\def\>{\rangle}
\def\<{\langle}
\def\width4{2.2in}
\def\figwidth1{2.5in}
\def\unitlen{2.3in}
\newcommand{\ket}[1]{{|#1\rangle}}
\newcommand{\bra}[1]{{\langle#1|}}
\twocolumn[\hsize\textwidth\columnwidth\hsize\csname 
@twocolumnfalse\endcsname 

\title{Vortex formation in two dimensions: When symmetry breaks, how
big are the pieces? }
\author{Andrew Yates$^{1,2}$ and Wojciech H. Zurek$^1$}
\address{$^{(1)}$Theoretical Astrophysics T-6, MS B-288; Los Alamos
National Laboratory, Los Alamos, NM 87455\\$^{(2)}$D\'{e}pt. de Physique
Th\'{e}orique, Universit\'{e} de Gen\`{e}ve, 24, quai E. Ansermet,
CH-1211 Gen\`{e}ve 4} 
\date{\today}
\maketitle

\begin{abstract} 
We investigate the dynamics of second order phase transitions in two
dimensions, breaking a gauged U(1) symmetry. Using numerical
simulations, 
we show that the density of topological defects formed scales with 
the quench timescale $\tau_Q$ as $n \sim \tau_Q^{-1/2}$ when the 
dynamics is overdamped at the instant when the freezeout of thermal
fluctuations takes place, and $n \sim \tau_Q^{-2/3}$ in the
underdamped case. This is predicted by the scenario proposed by one of
us \cite{WHZ1}. 
\end{abstract}
]
Phase transitions occur at all energy scales, from Bose-Einstein condensation 
near absolute zero to the sub-Planck temperatures relevant for symmetry 
breaking in a cosmological setting. While the dynamics of some types of 
first-order transitions ({\it i.e.}, the process of nucleation) has been 
extensively studied and is well understood for some time, analogous
understanding of second-order transitions is emerging only now.
In this case there is no supercooling and the final state of the system 
is asymptotically approached through phase ordering. Untill recently, research 
has focused largely on the asymptotic scalings of this post-transition
ordering \cite{Bray} rather than on the dynamics of the transition itself. 

The change in focus is relatively recent. Kibble, in a seminal paper
\cite{Tom76}, pointed out that topological defects may have
significant cosmological consequences -- {\it i.e.} they may act as
seeds for structure or as constraints on models. While
their distribution at the time they appear is largely forgotten by
the present, some of its features (such as the {\it ab initio}
existence of `infinite' string) are essential for the scenario. Moroever,
the initial density of topological defects may be directly relevant
for generation of baryons \cite{Perkins}.

The initial configuration of the order parameter established in the course of
the transition is therefore important. Furthermore, it is accessible in cases
where topological defects are formed, as they bear witness to the dynamics of 
the order parameter in the immediate vicinity of the critical point. Experiments
based on this idea allow one to probe the critical dynamics of symmetry 
breaking, and have been carried out in liquid crystals \cite{Yurke} and in
superfluids \cite{Lancaster,Helsinki,Grenoble}. This has already led to 
new insights into the dynamics of the transition between 
phases A and B of $^3$He \cite{Bunkov}.

We study the consequences of second
order phase transformation of the order parameter $\psi$, a complex
scalar field, with Landau-Ginzburg dynamics in two spatial dimensions
and the associated gauge field $A_a$. This is the Abelian
Higgs model coupled to a heat bath and with a dissipative term.
When cast into first order form, the equations of motion are the following
($\psi_X = (\psi_1, \psi_2)$, $i$
runs over $\{ x,y \}$, and $D_a \psi = \partial_a + ie A_a \psi$);
\bea
\pi_X & =& \partial_t \psi_X, \qquad
\Pi_i  = \partial_t A_i\\
\partial_t \pi_X & = & \nabla^2 \psi_X - e^2 A^2 \psi_X-
2e\epsilon_{XY} A^i \partial_i \psi_Y \nonumber \\
 &  & - \frac{\partial
V}{\partial \psi_X} - \eta\pi_X + \vartheta \label{eqn:scalar_eom}\\
\partial_t \Pi_i & = & \nabla^2 A_i - e\epsilon_{XY} \psi_X \partial_i
\psi_Y - e^2A_i |\psi|^2\\
V(\psi) & = & -{1\over 2} \epsilon m_0^2 \psi^2 + {1\over 4} \psi^4. 
\eea
The system is subjected to the white noise $\vartheta(x,y,t)$;
$
\langle \vartheta(x,y,t) \vartheta(x',y',t')\rangle = 2\eta\theta
\delta(x-x')\delta(y-y') \delta(t-t').
$
Here $\theta$ is the heat bath temperature and $\eta$ sets the rate at
which the field is damped, in accord with the fluctuation-dissipation theorem.

We induce the symmetry breaking by changing the sign of the dimensionless
parameter $\epsilon$ in the effective potential over the quench timescale
$\tau_Q$, so that 
$\epsilon = t/\tau_Q$ ($|\epsilon| \leq 1$).
The phase transition occurs when it becomes energetically and
entropically favourable for the order parameter to assume (in
equilibrium) a finite expectation value. In our case this happens in the
region $0 < \epsilon \ll 1$. The shift of critical temperature from
$\epsilon = 0$ occurs as a consequence of the coupling to the gauge
field ($m^2_{{\rm eff}} \sim m_0^2 + e^2 \langle A^2 \rangle $) and as a
result of finite temperature $\theta$, which we take to be $0.01$.

The equations of motion are evolved on a square lattice
of $512^2$ grid points,  using the standard staggered
leapfrog method with periodic boundary conditions. We impose the gauge
$A_t = 0$, implicit in Eqns. (2) and (3), and average each
case over 20 realisations. Defects were resolved by several grid
spacings at low temperatures.

We begin the simulations well above the phase transition, allowing the system 
to come to equilibrium under the influence of the noise and relaxation at 
constant $\epsilon$. We monitor the behaviour of the 
order parameter and the gauge field throughout the subsequent evolution.
The focus of attention, however, is the number of topological defects,
which can be identified as zeroes of $\psi$ in the broken symmetry phase.
Initially, in the symmetric phase, such zeroes are plentiful and short lived 
(see Fig. 1). While they cannot be identified with defects, their density 
and arrangement gives an idea of the nature of the thermal fluctuations.
\onecolumn
\twocolumn[\hsize\textwidth\columnwidth\hsize\csname 
@twocolumnfalse\endcsname
\begin{figure*}
\begin{minipage}{5in}
\setlength{\unitlength}{\unitlen}
\begin{picture}(3,1.20)(0,0)
\put(-0.05,0.20){\epsfxsize=\figwidth1\epsfbox{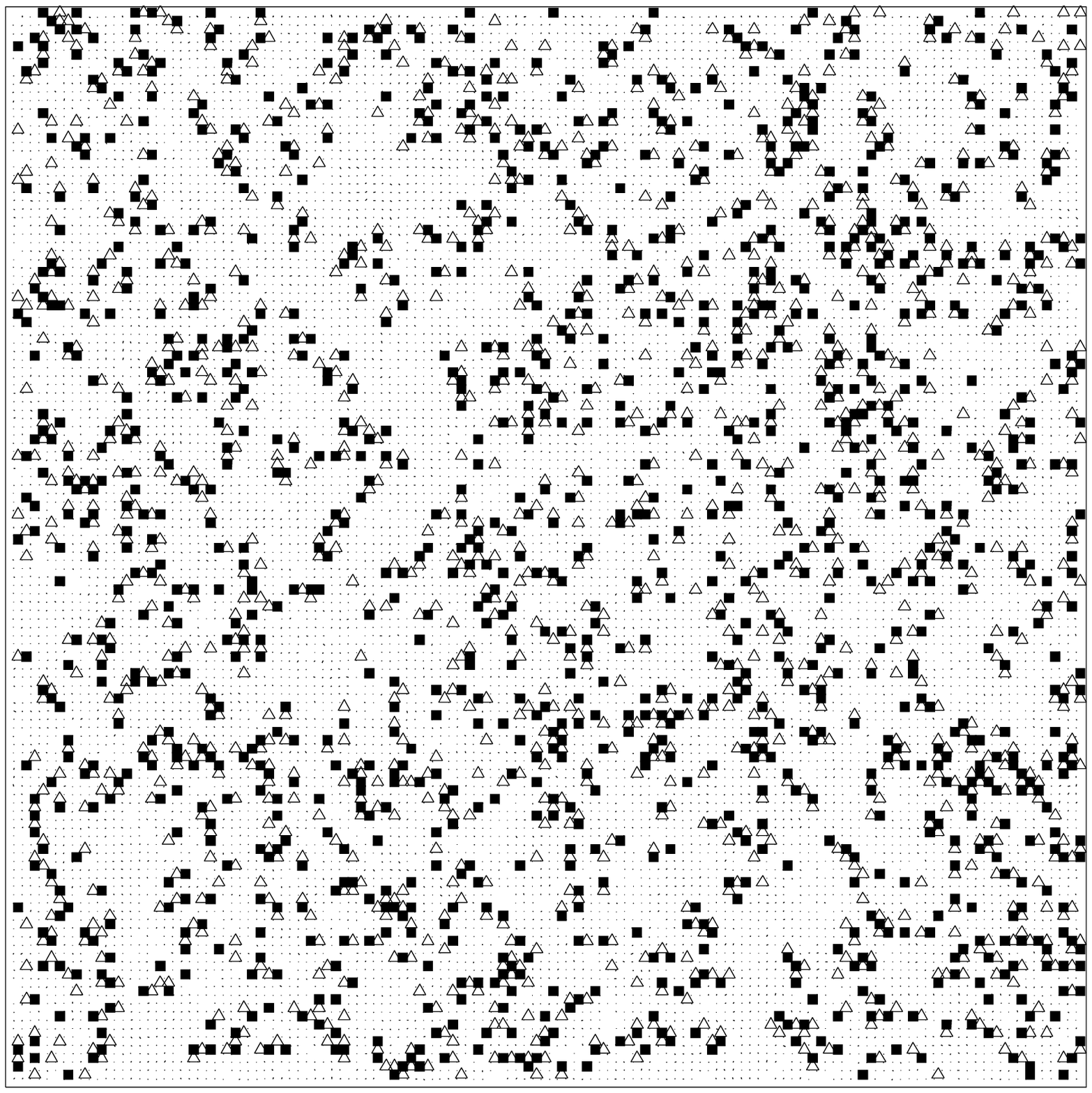}}
\put(0.95,0.20){\epsfxsize=\figwidth1\epsfbox{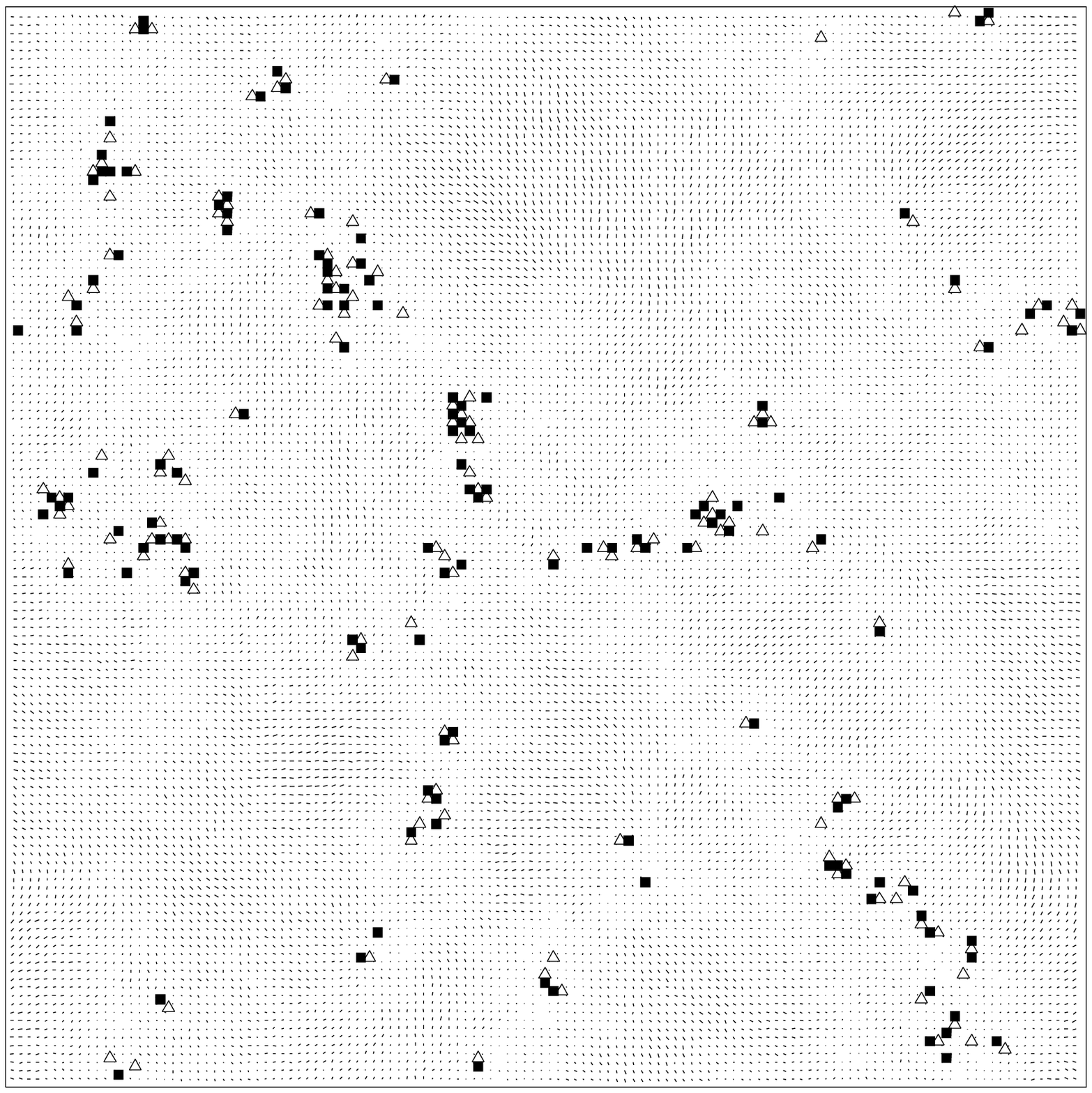}}
\put(1.95,0.20){\epsfxsize=\figwidth1\epsfbox{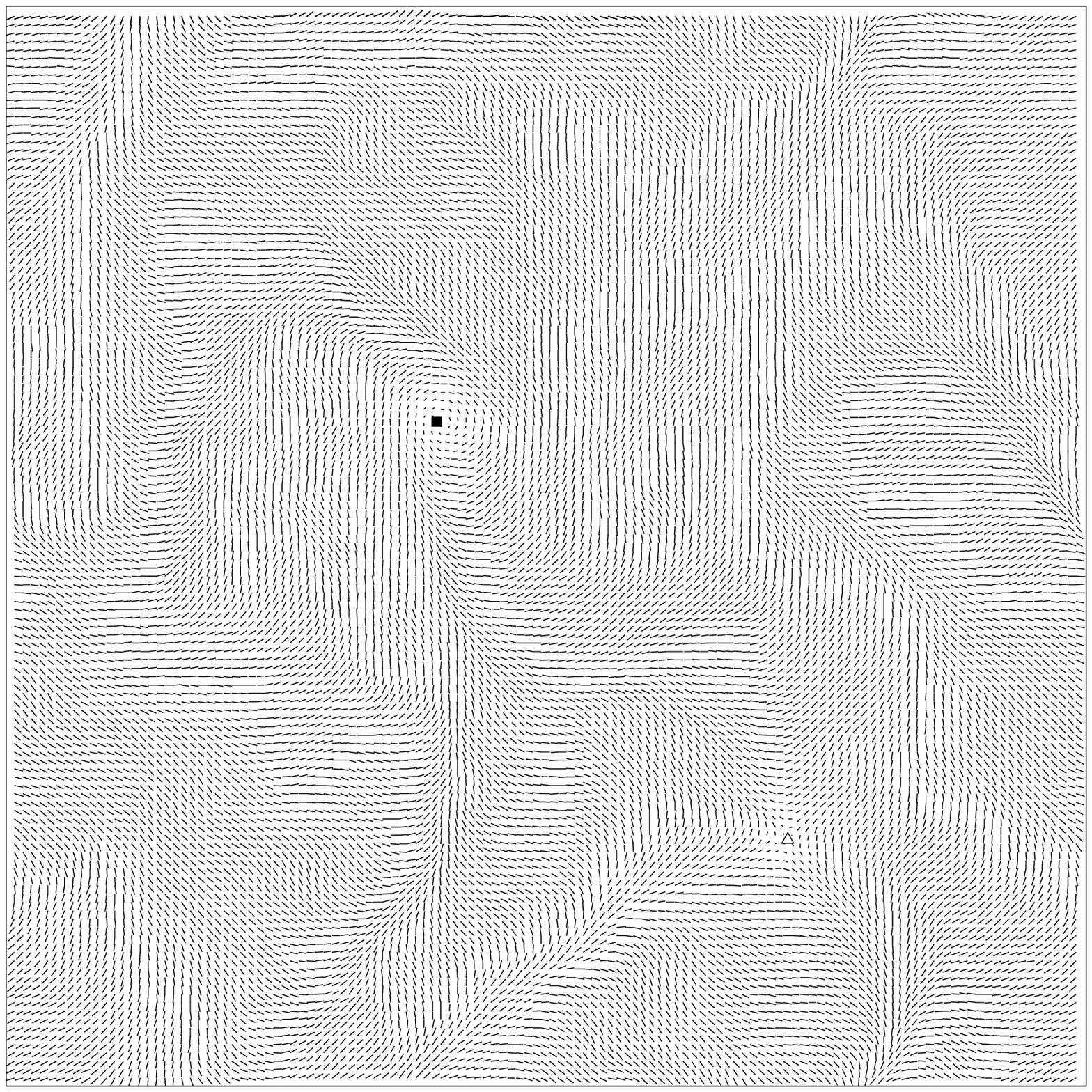}}
\put(0.5,0.25){(a)}
\put(1.5,0.25){(b)}
\put(2.5,0.25){(c)}
\put(0,0){\parbox[b]{7in}{$\;$ FIG. 1.
Representations of the field and defects (filled squares or open
triangles) on $128^2$ grid during and after a quench of
timescale $\tau_Q=32$, with $\eta=1.0$
and in the presence of a gauge 
field ($e=0.5$). Figure (a) shows the field at the critical point
($\epsilon=0$), (b) at $\epsilon=0.6$ and 
(c) at $\epsilon=1$, time $t=2\tau_Q$ after the critical point.
}}
\end{picture}
\end{minipage}
\label{fig:evol}
\end{figure*}
] 
Below the transition their overall density decreases, although 
there still exist regions where the field has a near-zero expectation
value and, 
hence, plenty of unstable zeros. Eventually, only a few isolated 
defects (as well as a few pairs which are about to annihilate) remain.
We count only zeroes that have no companions within $\xi_0 = m_0^{-1}$.

The local and global ($e=0$) gauge cases are qualitatively
indistinguishable until this late stage. 
However, local defects do not interact significantly over more than a
few correlation lengths and annihilate relatively
slowly, even when the friction coefficient is small. By contrast,
global defects interact via a potential logarithmic in distance, and
disappear much more rapidly. Estimates of initial defect densities
become more difficult in this case. Below, we shall focus primarily on
the local case, leaving the detailed global-local comparisons for the
future.

The theory of defect formation combines the realisation, due to Kibble
\cite{Tom76}, that the domains of the order parameter $\psi$ which
break symmetry incoherently must contain of the order one `fragment'
of a defect, with the estimate \cite{WHZ1} of the relevant size based
on the comparison of the relaxation timescale of $\psi$ with the
effective rate of change of the mass parameter $\epsilon$. In the
immediate vicinity of the critical temperature, the dynamics of $\psi$
are subject to critical slowing down. Thus, the timescale
$\tau$ over which the order parameter can react is given by
\be
\tau_{\dot \psi} = \frac{\eta \tau_0^2}{|\epsilon|}, \qquad \tau_{\ddot \psi} =
\frac{\tau_0}{|\epsilon|^{1/2}}, \qquad (\tau_0= 1/m_0)
\ee
in the overdamped and  underdamped cases respectively,
where correspondingly the first or second time derivative in
Eqn. \ref{eqn:scalar_eom} dominates.  The overdamped scenario is
presumably more 
relevant for condensed matter applications, while in the cosmological settings 
$\psi$ may be underdamped.

The characteristic timescale of variations of $\epsilon$ is
$\epsilon / {\dot \epsilon} = t$. The system is able to
readjust to the new equilibrium as 
long as the relaxation time is smaller than $t$. Hence, outside the
time interval $[- \hat t,\hat t\,]$ defined by the equation
$\tau(\epsilon(\hat t)) = \hat t$,
the evolution of $\psi$ is approximately adiabatic, and physical
quantities follow their equilibrium expectation values. Thus the times
\be
\hat{t}_{\dot \psi} = \pm \tau_0 \sqrt{\eta \tau_Q}; \qquad
\hat{\epsilon}_{\dot \psi} = \pm \sqrt{\frac{\eta \tau_0^2}{\tau_Q}}
\ee
and
\be
\hat{t}_{\ddot \psi} = \pm m_0^{-2/3} \tau_Q^{1/3}; \qquad
\hat{\epsilon}_{\ddot \psi} = \pm \left( \frac{m_0^2}{\tau_Q} \right)^{2/3}
\ee
mark the borders between the (approximately) adiabatic and impulse
(or, perhaps,``drift'') stages of evolution of $\psi$ in the
overdamped and underdamped  cases, respectively.

In particular,  the correlation length $\xi$ of $\psi$ above the
transition will cease to increase with the Landau-Ginzburg scaling
($\xi = \xi_0 /|\epsilon|^{1/2}$)
above the transition once the adiabatic--impulse boundary at $-\hat
t$ is reached. Dynamics will be suspended (except for the drift and noise)
 in the interval
$[- \hat t,\hat t\,]$ and will resume at $+\hat t$ below the
transition, as the mass term drives the symmetry
breaking.

We expect, then, that the characteristic length scale over which
$\psi$ is ordered already in the course of the transition will be the
correlation length at freeze-out, $\hat \xi = \xi_0 / \sqrt{\hat
\epsilon}$ \cite{WHZ1}.  This results in
\be
\hat{\xi}_{\dot \psi} =
\xi_0 \left (\frac{\tau_Q}{\eta \tau_0^2} \right)^{1/4}, \;\;\;
\hat{\xi}_{\ddot \psi} =
\xi_0 \left (\frac{\tau_Q}{\tau_0}\right)^{1/3}
\ee
in the two cases. The initial density of vortex lines in two
dimensions should then scale as
\bea
n_{\dot \psi} &=&  \frac{1}{(f_{\dot \psi}\hat{\xi}_{\dot \psi})^2} =
\frac{1}{(f_{\dot \psi} \xi_0)^2} \sqrt{\frac{\eta \tau_0^2}{\tau_Q}},
\label{eqn:od_dens}\\ 
n_{\ddot \psi} &=& \frac{1}{(f_{\ddot \psi}\hat{\xi}_{\ddot \psi})^2}
=\frac{1}{(f_{\ddot \psi} \xi_0)^2} \left( \frac{\tau_0}{\tau_Q}
\right)^{2/3}
\label{eqn:ud_dens} 
\eea
in the over- and under-damped regimes respectively. Above, $f$ is the 
proportionality factor which is expected to be of the order of a few 
\cite{PabWoj}, and which may be estimated analytically in some cases
\cite{Gill}. 

The relative importance of the $\dot \psi$ and $\ddot \psi$ terms in
Eqn. (\ref{eqn:scalar_eom}) also depends on $\hat \epsilon$. What matters
for the formation of defects is -- in view of the arguments above
-- which of the two terms dominates at $\hat t$. Thus, $\eta^2 > \hat
\epsilon$ is the condition for overdamped evolution, which leads to
the inequality $\eta^3 \tau_Q > m_0^2$.
Therefore, one can enter the overdamped regime by performing a sufficiently
slow quench, as well as increasing the damping parameter $\eta$.

We verify these scalings in Fig. 2, where the numbers of defects obtained in
both the overdamped and underdamped cases are plotted as a function of 
the quench time $\tau_Q$. In the overdamped case the annihilation of the 
well-separated topological defects is slow. Consequently, it is relatively 
easy to count them at some fixed time. This yields Fig. 2b with 
$n \sim \tau_Q^{\gamma}$, $\gamma=-0.44 \pm 0.1$ for $\tau_Q \ge 8$, 
in good agreement with the theoretical 
prediction of $\gamma=-1/2$ \cite{WHZ1}. A similar
conclusion can be  
reached for the underdamped case, Fig. 2a, except that annihilation is now 
quite rapid, and the initial number of defects is harder to define. 
This is especially true for the fastest quenches which produce most defects.
However, when the two left-most points most affected by annihilation 
are ignored, straightforward counting of vortices yields 
$\gamma= -0.6\pm 0.07$. This is in agreement with the theory -- which 
in this case predicts $\gamma= -2/3$ \cite{WHZ1,WHZ2} -- and with
the indications from kink formation in one dimension \cite{LagZur}.  

The reason the slope may be expected to become more shallow for small $\tau_Q$
is easy to understand. A very fast quench becomes indistinguishable
from an instantaneous one, which does not allow for the adiabatic regime
we have noted above. Instantaneous quenches start the evolution in the 
broken symmetry phase, but with an initial field configuration which
will contain many zeros per healing-length size volume -- too many to regard 
them as well-defined defects. Unless the defects are
well-separated at their  
conception (which in effect requires $\hat \epsilon \ll 1$), annihilation will 
decide their initial density. We see this already in Fig. 2b, where 
$\hat \epsilon = 0.5$ for the left-most point. The effect is even more 
pronounced in the underdamped case, which allows for more rapid annihilation
(Fig. 2a). 

The annihilation rate can be expected to be of the simple form
$\dot n = - \chi n^2$,
where $ \chi$ is a function of $\eta$, as
the annihilation is proportional to the rate at which defects
encounter one another.
This leads to
\be
n^{-1} = n_0^{-1} + \chi t.
\label{eqn:anni_soln}
\ee
Fig. 3 shows (in the
underdamped case, when the annihilation is appreciable) the fit
between (\ref{eqn:anni_soln}) and the data.
We can then infer the value of the initial defect density $n_0$, which is also
shown in Fig. 2a. This procedure (followed, for instance, 
in the superfluid work, \cite{Lancaster}) yields a somewhat different slope
$n_0 \sim \tau_Q^{-0.79 \pm 0.04}$, steeper than equal-time data, and somewhat
steeper than the theoretical prediction. It also seems to succesfully correct
for the annihilation (although the left-most point still seems to be affected).
Also, using the data in Fig. 2,
we estimate $f_{\dot \psi} \simeq 12$ and $f_{\ddot \psi} \simeq 8$
(equal-time measurements), $f_{\ddot \psi} \simeq 4$ (from fitting to
Eqn. \ref{eqn:anni_soln}).

One of the issues in the formation of topological defects is the relative 
significance of the thermal fluctuations, which continue to re-arrange the 
configuration of the order parameter down to the Ginzburg regime ({\it
i.e.} 
below the phase transition temperature). If, as was once thought, fluctuations 
and Ginzburg temperature were to determine initial defect density, then taking 
the system above the critical point would be expected to erase the
pre-existing  
configuration of defects and create a new one. We have performed a numerical 
experiment designed to test the importance of fluctuations. The initial 
configuration with defects present is the one shown above in
Fig. 1c. We reheat this by varying
$\epsilon$ from the broken symmetry value $\epsilon = 1$ to several
values in the vicinity of the critical point $\epsilon\simeq0$, using
the same $\tau_Q$ was used as in the original quench of Fig. 1. 

The results (Fig. 4) demonstrate that even when the system is taken above the
critical point, the initial configuration of defects is eventually recovered,
as long at the re-heating does not take it further than $\hat \epsilon$ into
the symmetric phase. This confirms the theory put forward in \cite{WHZ1}, and
leads one to conclude that thermal fluctuations cannot significantly re-arrange
configurations of the order parameter on scales larger than $\sim \hat \xi$,
unless the ``impulse strip'' $|\epsilon| < |\hat \epsilon|$ is traversed
(or unless the time spent in that regime is set by a timescale other than
the original quench timescale $\tau_Q$).

We have used high-resolution numerical simulations to explore phase
transitions in two dimensions and have found the scaling with quench
timescale and damping agree with the predictions of the
Kibble-Zurek scenario. The importance of the freeze-out time $\hat t$
as the defining moment for defect formation has been illustrated. 

The authors would like to thank Pablo Laguna for the use of his code
and Mike Warren for assistance with the Loki parallel computing
facility at Los Alamos.

\onecolumn
\begin{figure*}
\begin{minipage}{4.5in}
\setlength{\unitlength}{\unitlen}
\begin{picture}(3,1)(0.1,0.2)
\put(-0.05,0.1){\epsfxsize=2.4in\epsfbox{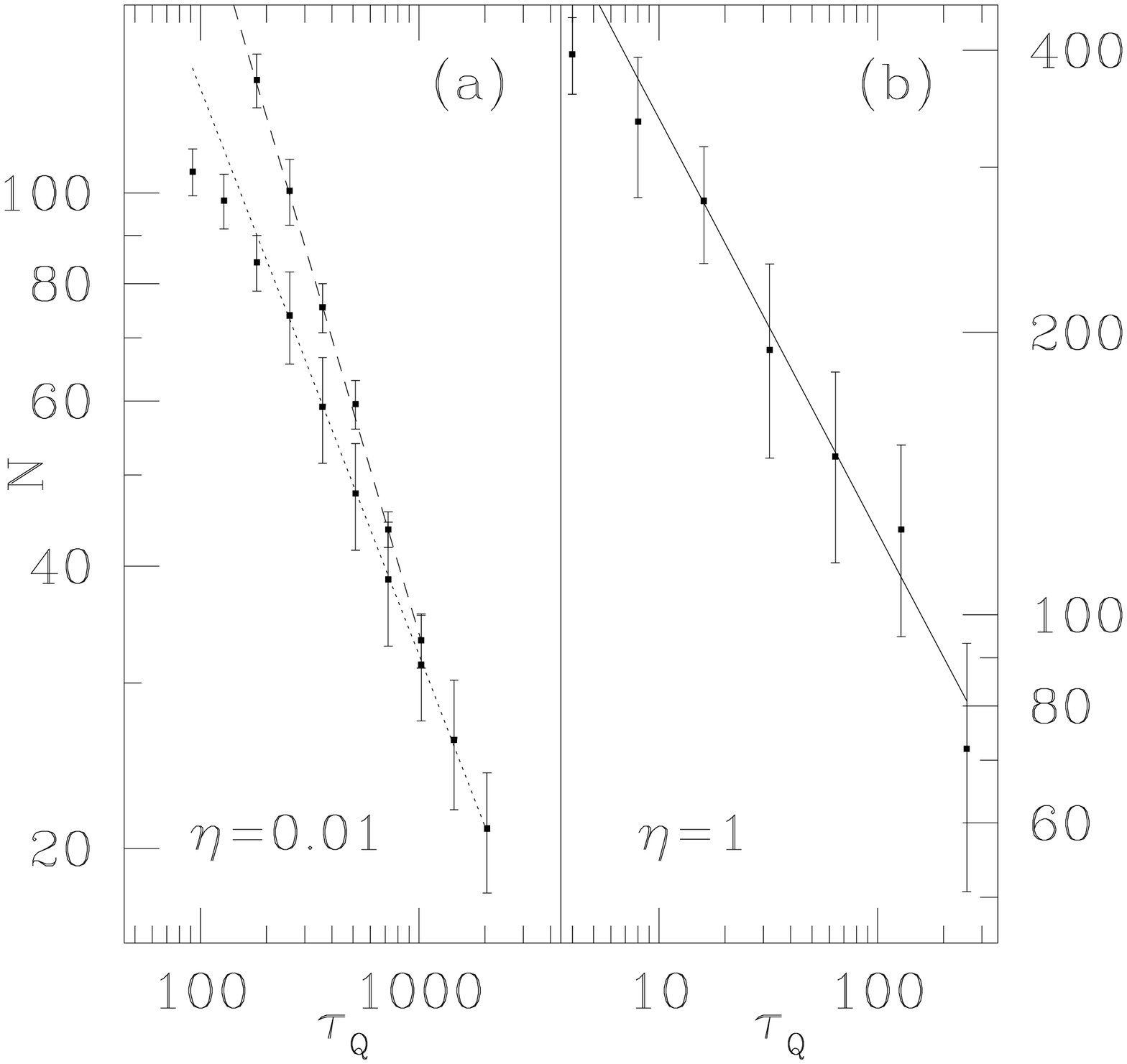}}
\put(2.1,0.1){\epsfxsize=2.4in\epsfbox{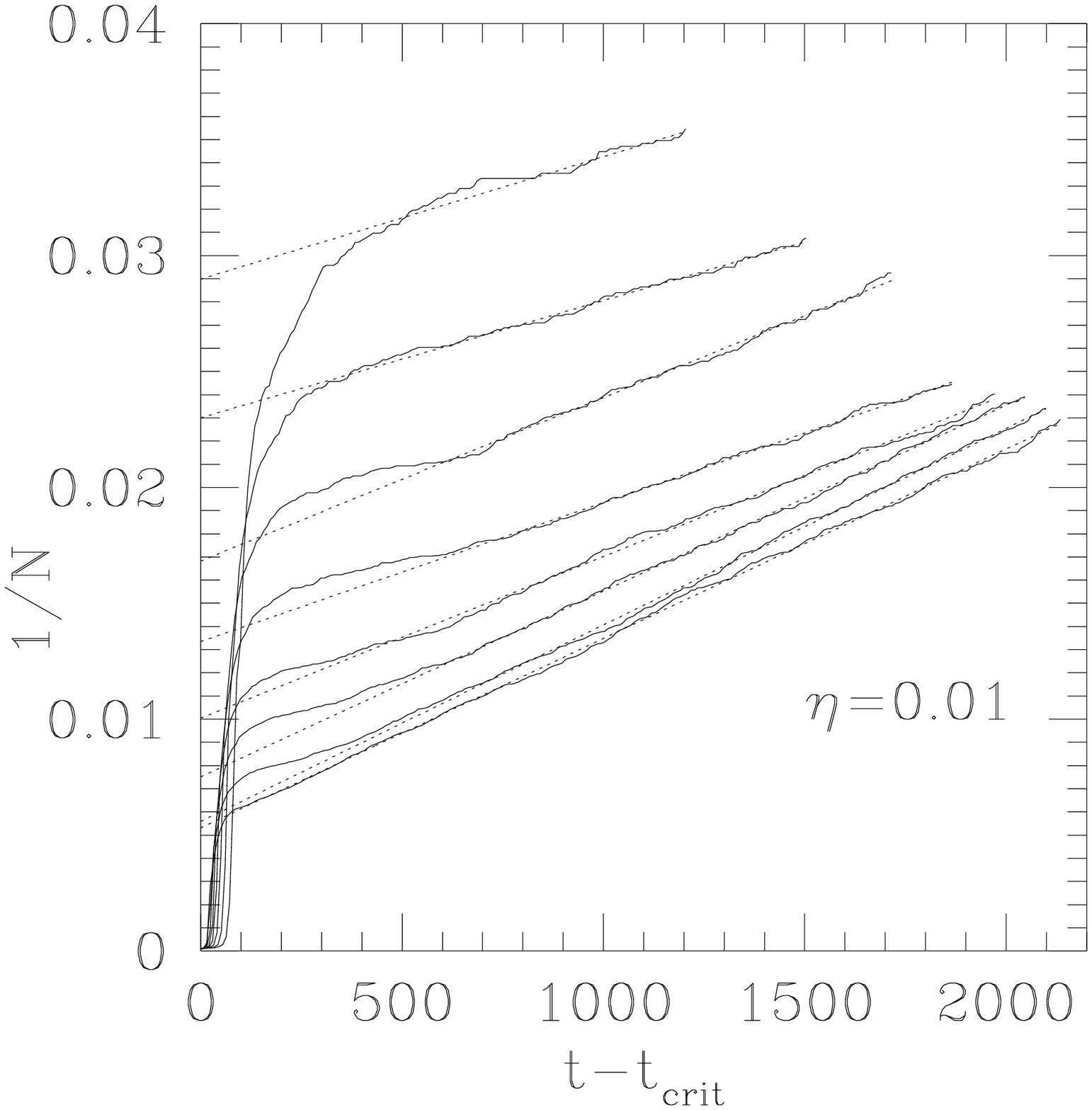}}
\put(1.03,0.1){\parbox[b]{2.4in}
{FIG. 2 (left)
The variation of defect count $N$ with
$\tau_Q$ in both (a) under- and (b) over-damped regimes.
The fitted slopes are (a) $-0.60 \pm 0.07$ ($\chi^2 = 0.034$, dropping
the 3 left-most points) 
 and (b) $-0.44 \pm 0.10$ ($\chi^2 = 0.44$, dropping 2 points)
Predicted values were $-2/3$ and  $-1/2$. The points fitted
by the dashed line are inferred from the fits in
fig. 3. The slope is $-0.79 \pm 0.04$ ($\chi^2 = 0.27$).\\[0.1cm]
$\;$ FIG. 3 (right)
Defect annihilation in the underdamped
regime. The inverse of the defect count is plotted against time for
various $\tau_Q$. Also shown are least-squares fits using
Eqn. \ref{eqn:anni_soln}.}}
\end{picture}
\end{minipage}
\end{figure*}
\begin{figure*}
\begin{minipage}{5in}
\setlength{\unitlength}{\unitlen}
\begin{picture}(3,2.75)(0.15,0)
\put(0.01,1.8){\epsfxsize=\width4\epsfbox{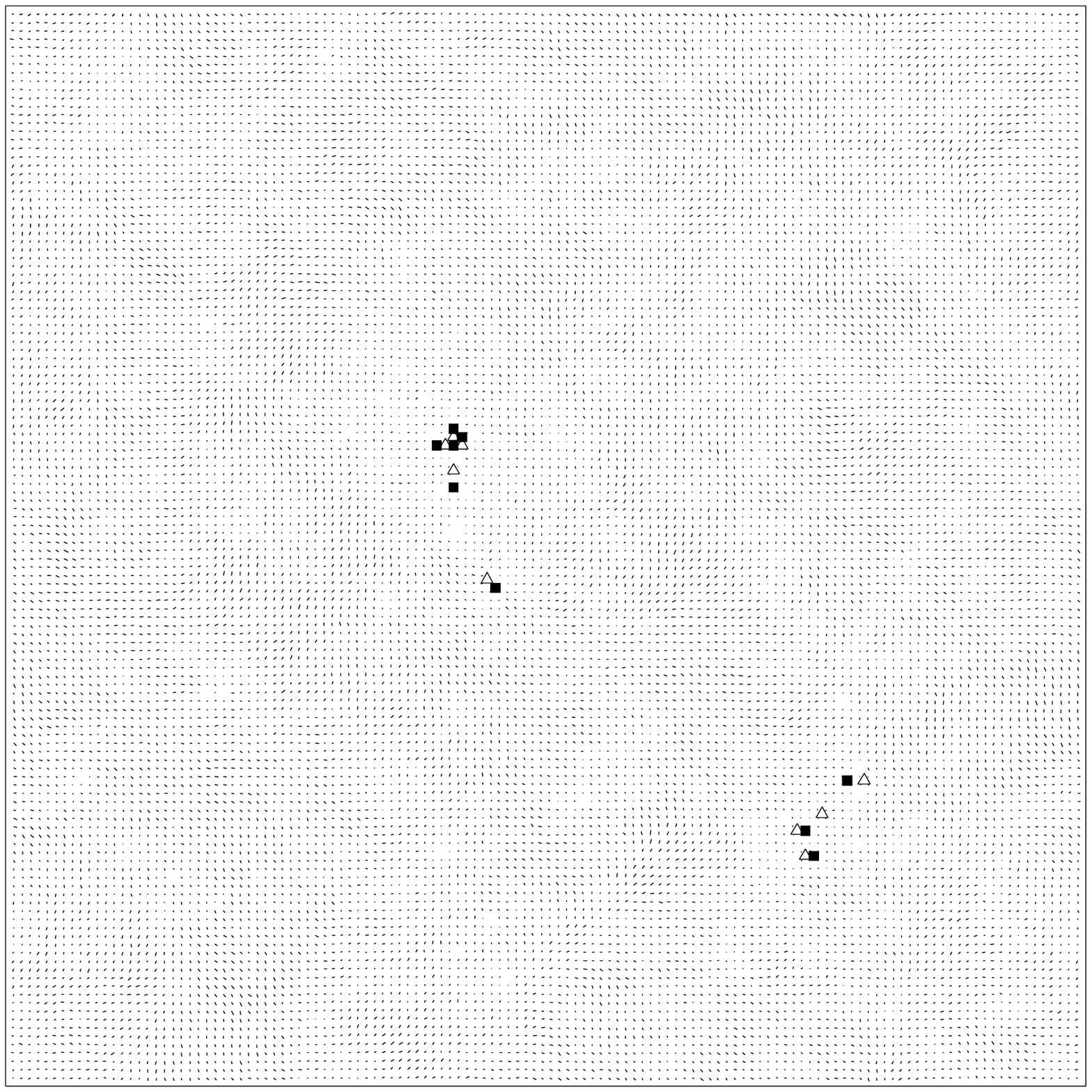}}
\put(0.01,0.95){\epsfxsize=\width4\epsfbox{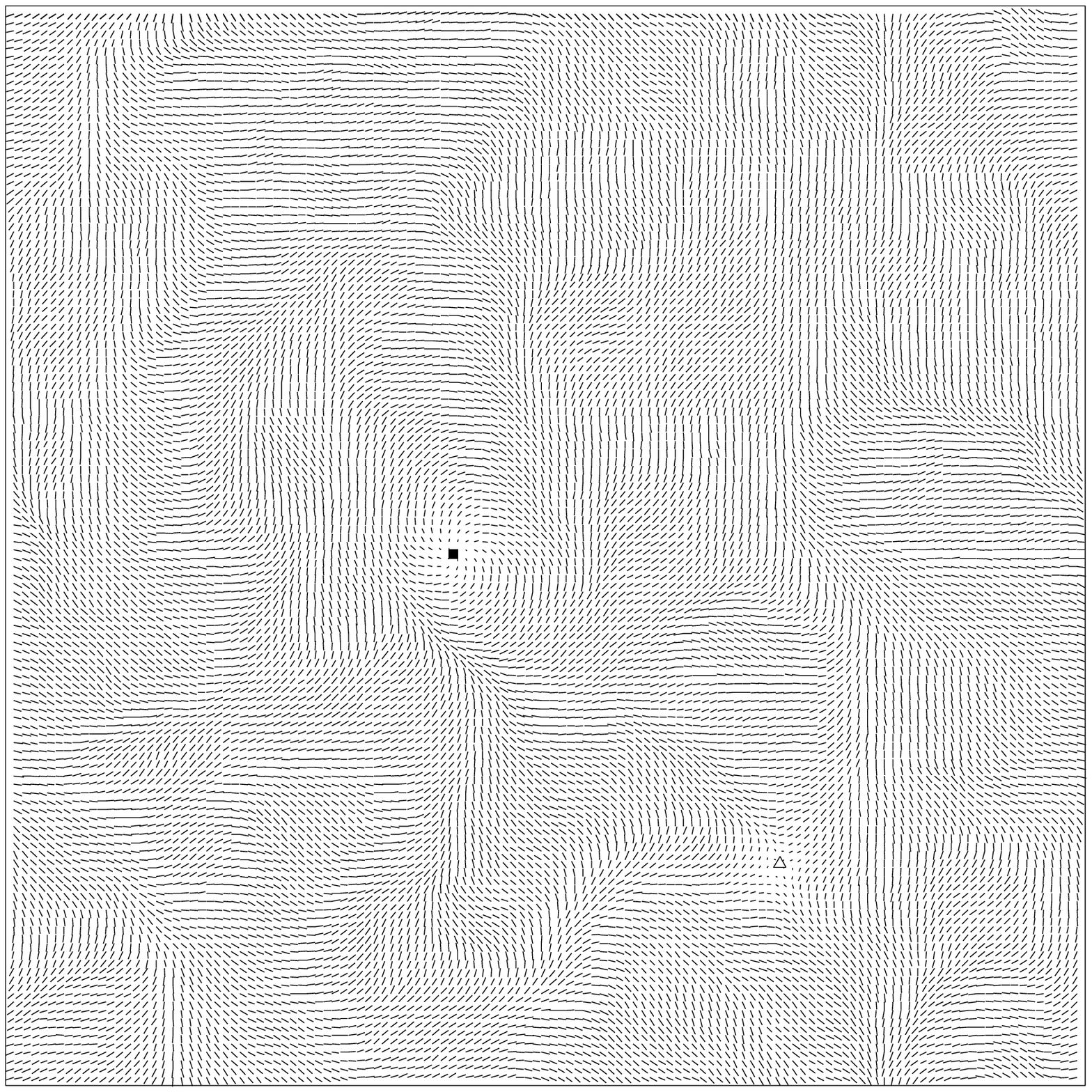}}
\put(1.05,1.8){\epsfxsize=\width4\epsfbox{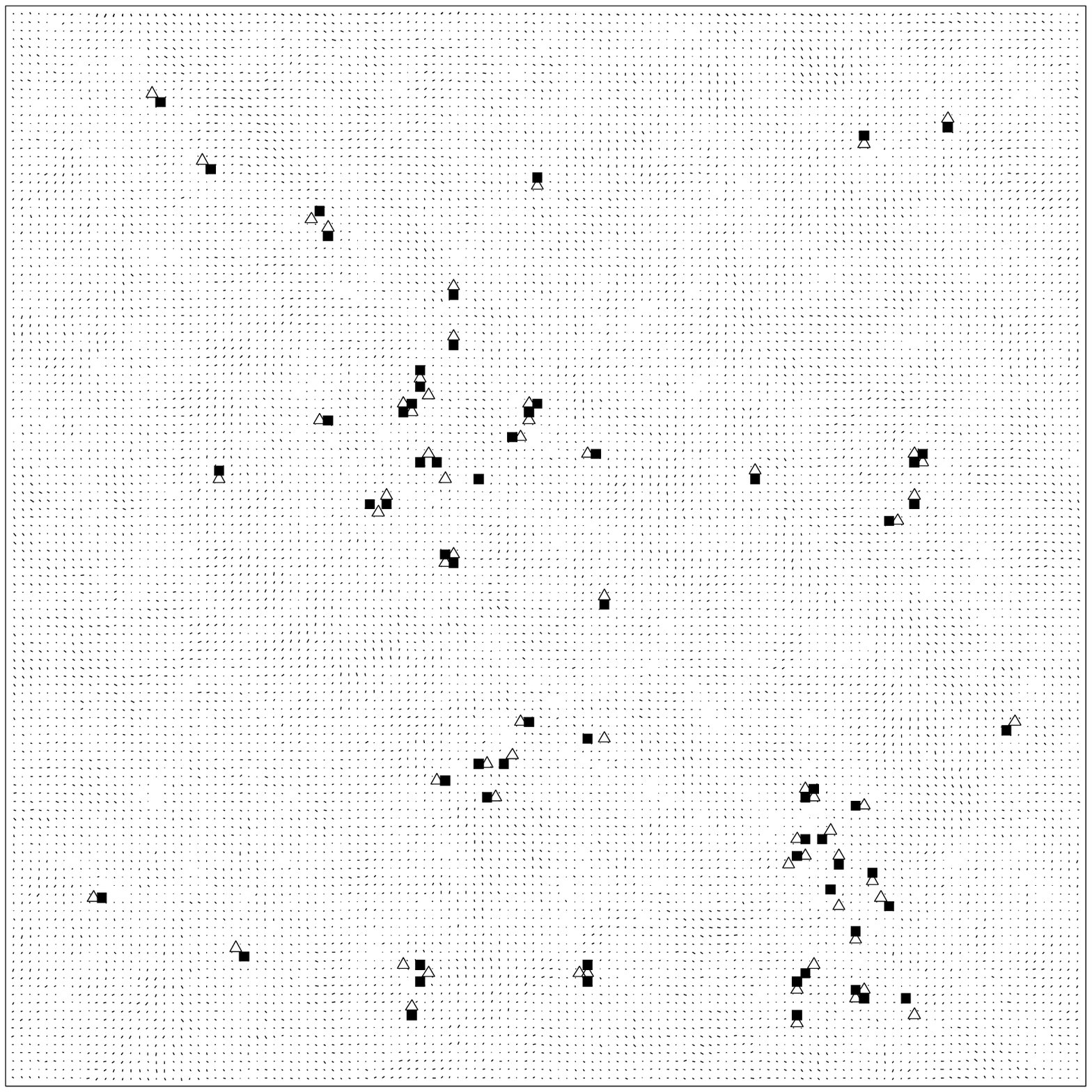}}
\put(1.05,0.95){\epsfxsize=\width4\epsfbox{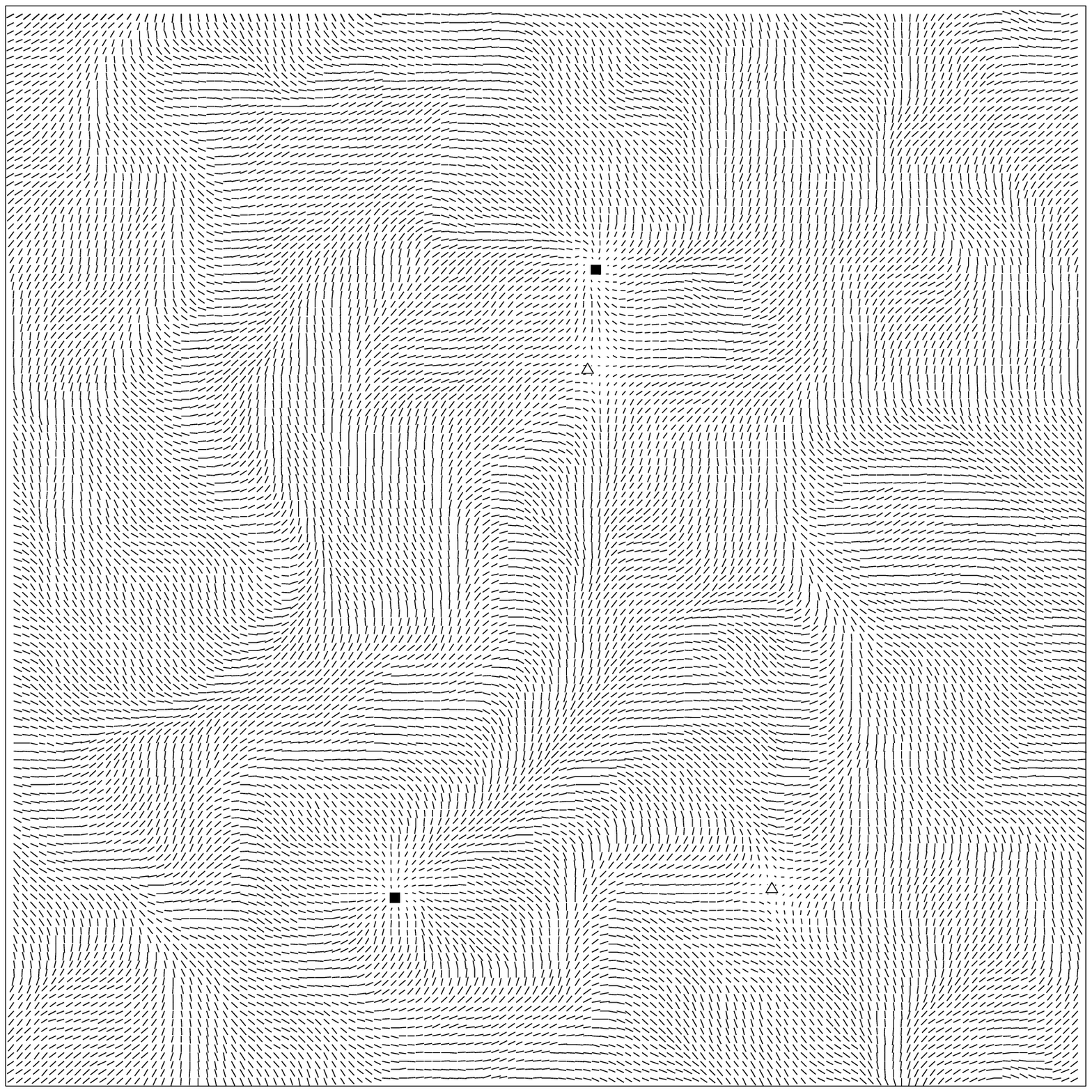}}
\put(2.09,1.8){\epsfxsize=\width4\epsfbox{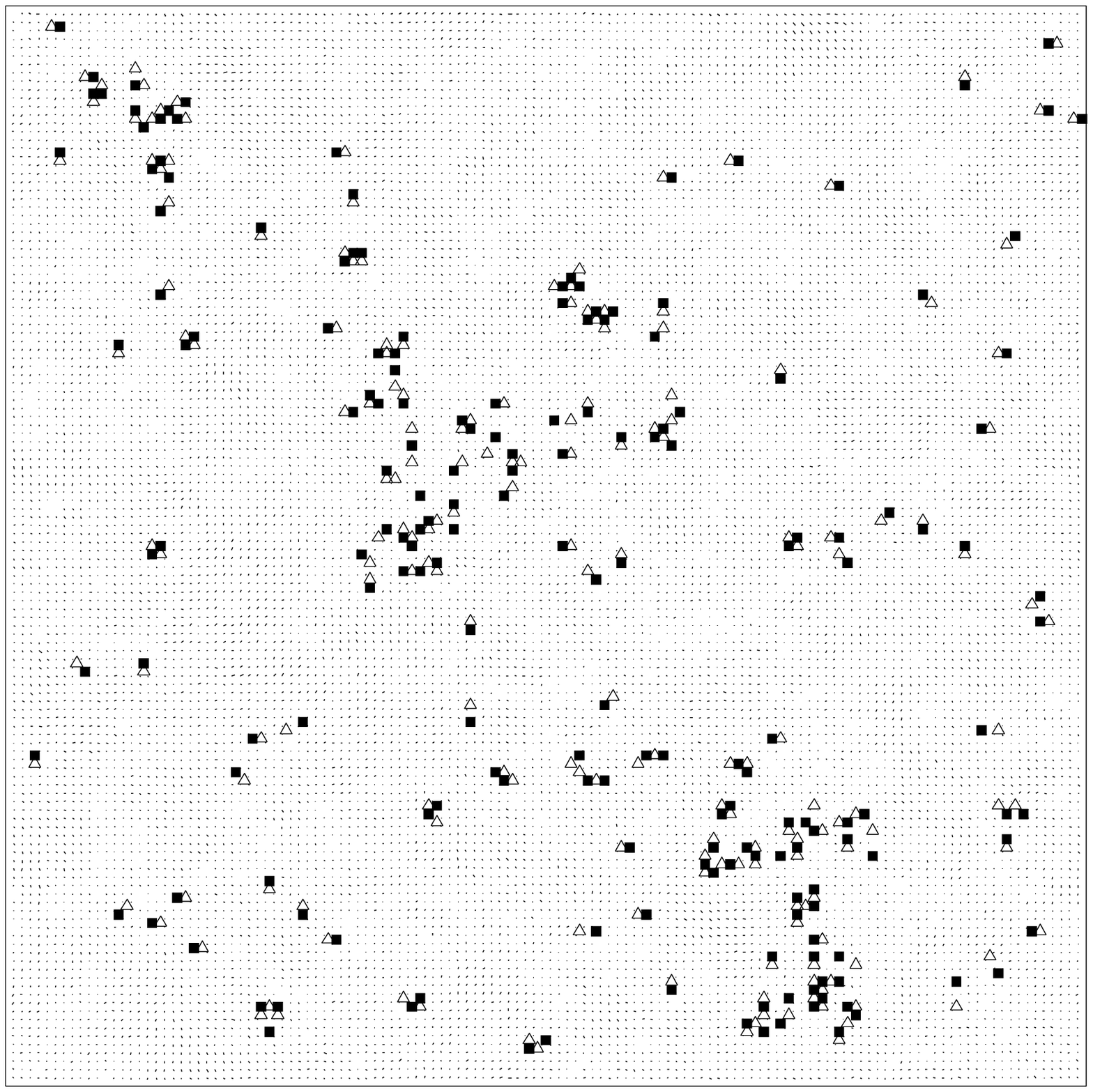}}
\put(2.09,0.95){\epsfxsize=\width4\epsfbox{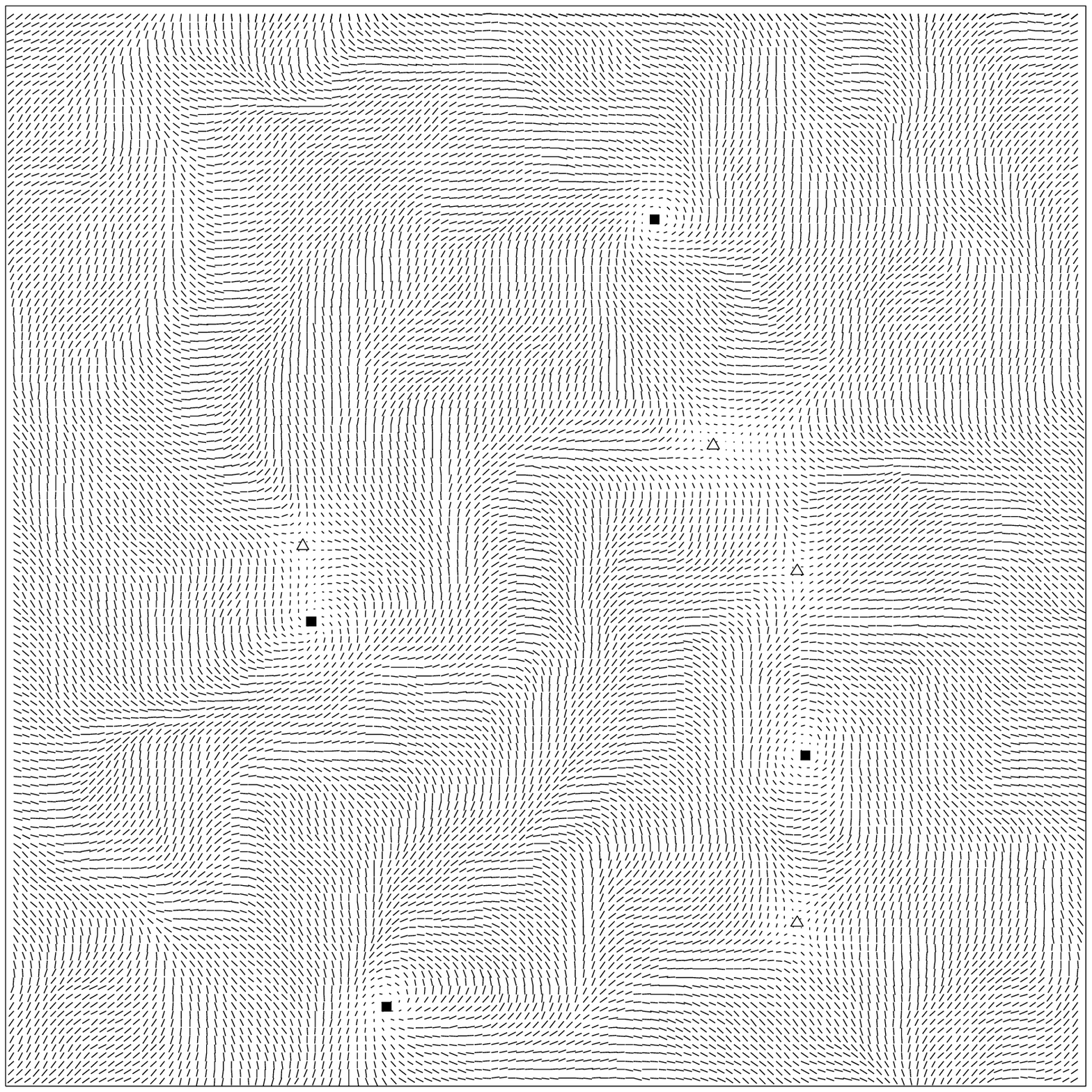}}
\put(0.0,0.47){\epsfxsize=2.4in\epsfbox{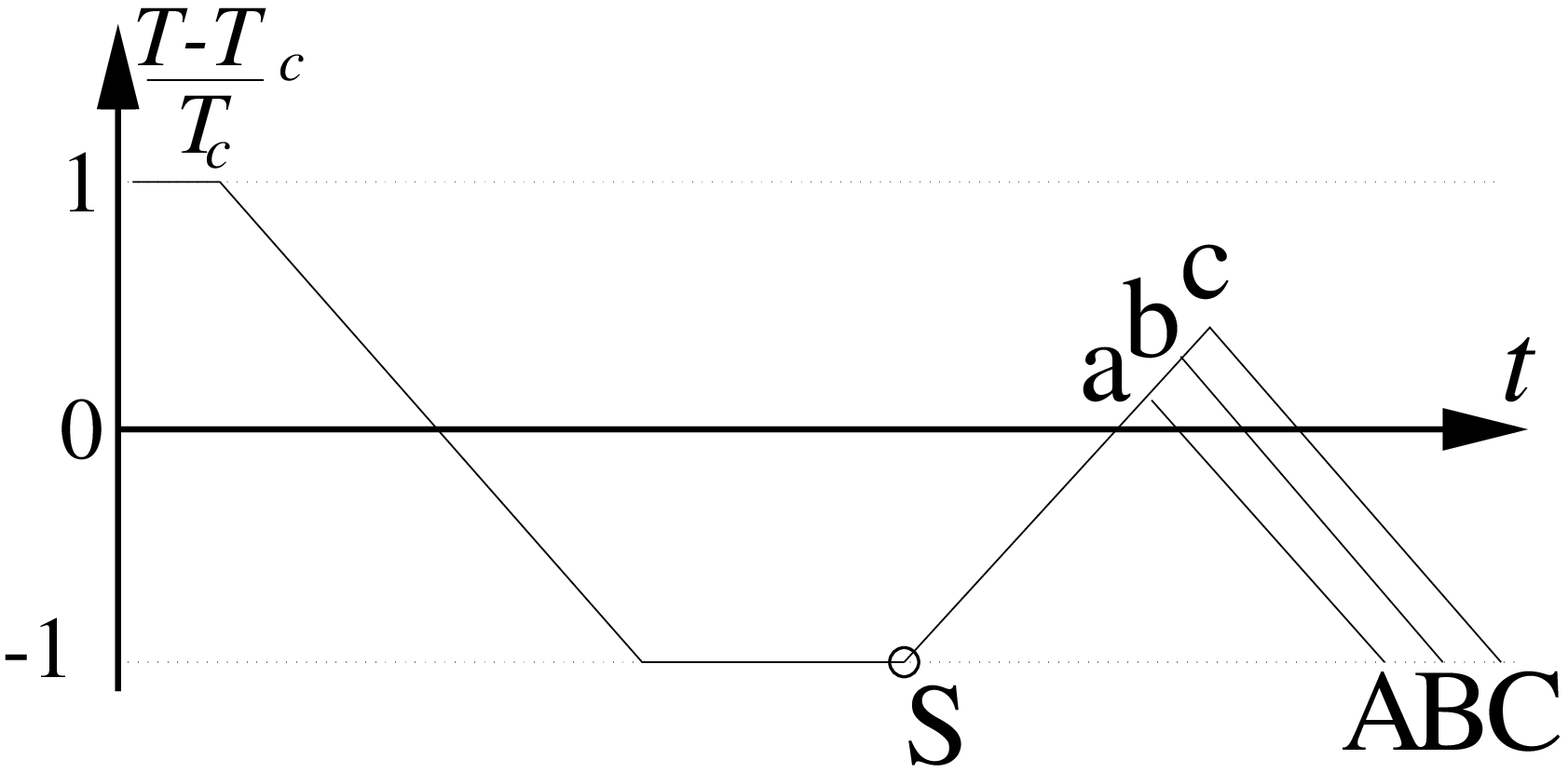}}
\put(1.1,0.95){\parbox[t]{4.8in}{$\;$ FIG. 4. Memory and
reheating. The post-transition configuration shown in figure 1
(the point S in the sketch to the left) is reheated to various
temperatures close to the critical point  
(figs. a-c) and cooled again to $\epsilon = 1$ (A-C
respectively).
The reheat values of $\epsilon$ are (a)
$-0.1$, (b) $-0.2$ and (c) $-0.25$. Here $|\hat \epsilon| \simeq 0.17$.
Notice that the memory of the configuration is largely preserved even 
when the critical temperature is exceeded during reheating (figs. A and B).
Memory is erased only when the temperature crosses the $|\hat
\epsilon|$ freezeout zone associated with the formation of the
original defect configuration.
}}
\put(0.94, 1.94){{\bf (a)}}
\put(0.94, 1.09){{\bf (A)}}
\put(1.98, 1.94){{\bf (b)}}
\put(1.98, 1.09){{\bf (B)}}
\put(3.02, 1.94){{\bf (c)}}
\put(3.02, 1.09){{\bf (C)}}
\end{picture}
\end{minipage}
\label{fig:reheating}
\end{figure*}
\end{document}